\documentclass[preprint,aps,floats,nofootinbib]{revtex4}
\input epsf.tex
\def\DESepsf(#1 width #2){\epsfxsize=#2 \epsfbox{#1}}

\newcommand{\no}{\nonumber\\}

\newcommand{\be}{\begin{equation}}
\newcommand{\ee}{\end{equation}}
\newcommand{\bea}{\begin{eqnarray}}
\newcommand{\eea}{\end{eqnarray}}

\newcommand{\areq}{~&\hspace{-2.35mm}=&\hspace{-2.35mm}~}

   \def\thebibliography#1{\centerline{\bf REFERENCES}
     \list{[\arabic{enumi}]}{\settowidth\labelwidth{[#1]}\leftmargin
     \labelwidth\advance\leftmargin\labelsep\usecounter{enumi}}
     \def\newblock{\hskip .11em plus .33em minus -.07em}\sloppy
     \clubpenalty4000\widowpenalty4000\sfcode`\.=1000\relax}

\begin{document}
\preprint{\vbox{\hbox{}\hbox{}\hbox{KEK-TH-876}}}

\vspace*{0.5cm}

\title{$B \to \eta^{(\prime)} l \nu$ decays
and the flavor-singlet form factors}

\author{ \vspace{0.5cm}
C.~ S. ~Kim$^a$\footnote{cskim@yonsei.ac.kr}, ~~
Sechul ~Oh$^b$\footnote{scoh@post.kek.jp} ~~ and ~~
Chaehyun ~Yu$^a$\footnote{chyu@cskim.yonsei.ac.kr}}

\address{ \vspace{0.3cm}
$^a$Department of Physics and IPAP, Yonsei University, Seoul
120-479, Korea \\
$^b$Theory Group, KEK, Tsukuba, Ibaraki 305-0801, Japan\\
\vspace{1cm}}

\draft

\vspace{1cm}

\begin{abstract}

\noindent
We study semileptonic decays $B \to \eta^{(\prime)} l \nu$,
taking into account the flavor-singlet contribution
($F^{\rm singlet}_+$) to the $B \to \eta^{(\prime)}$ form factors,
which arises from the two-gluon emission in a decaying $B$ meson.
It has been recently pointed out that, in addition to large weak
annihilation effects, the unknown value of $F^{\rm singlet}_+$
prevents accurate theoretical estimates in the analysis of
$B \to \eta^\prime K$ decays in QCD factorization. We present a
certain method to determine $F^{\rm singlet}_+$ with a reasonable
accuracy, using $B \to \eta^{(\prime)} l \nu$ and $B \to \pi l \nu$
decays.
We also investigate the possible effect of $F^{\rm singlet}_+$ on
the estimated branching ratios (BRs) for
$B \to \eta^{(\prime)} l \nu$ and find that the BR for
$B \to \eta' l \nu$ is particularly sensitive to the effect of
$F^{\rm singlet}_+$.
\end{abstract}

\maketitle

\newpage

Semileptonic decays of $B$ mesons have been extensively studied with particular
interests.
They can serve as useful applications of various non-perturbative theoretical
approaches and provide an efficient way for the determination of
the Cabibbo-Kobayashi-Maskawa (CKM) matrix elements, such as $V_{cb}$ and
$V_{ub}$.
In fact the present best experimental data for $V_{ub}$ come from measurements
of the exclusive semileptonic decays $B \to \pi l \nu$ and $B \to \rho l \nu$
(CLEO Collaboration \cite{CLEO1}),
and the inclusive semileptonic decay $B \to X_u l \nu$ (LEP Heavy Flavor
Group \cite{LEP}):
\begin{eqnarray}
  |V_{ub}| &=& ( 3.25 \pm 0.14^{+0.21}_{-0.29}
  \pm 0.55) \times 10^{-3} ~~ [{\rm CLEO}]~, \nonumber \\
           &=& ( 4.09^{+0.36+0.42}_{-0.39-0.47} \pm 0.25
  \pm 0.23) \times 10^{-3} ~~ [{\rm LEP}]~,
\end{eqnarray}
as well as the exclusive charmless nonleptonic
$B \to D_s \pi$ decay \cite{KKLN}.
Recently BELLE \cite{BELLE} and BABAR \cite{BABAR}, respectively,
have announced preliminary results for $|V_{ub}|$ that are similar
to the CLEO result.  In the analyses, they also used the exclusive
semileptonic processes: $B \to \pi l \nu$ (BELLE) and
$B \to \rho l \nu$ (BABAR).
Although these measurements currently suffer from large uncertainties
due to model-dependence, a dominant background, and so forth,
a more accurate value of $|V_{ub}|$ will become available through
future studies
by using the hadronic invariant mass of the inclusive decay \cite{kimbarger}.

An analysis of the charmless semileptonic $B$ decays
involves the non-perturbative hadronic form factors whose theoretical
estimation is usually model-dependent.  Over the past few years, there has been
considerable progress in the calculations of the $B \to \pi$ form
factor, based on various theoretical approaches, such as the lattice QCD
calculation \cite{lattice}, the light-cone QCD sum rule (LCSR) \cite{lightcone1},
and so on.  It is known that these approaches produce consistent results for
the $B \to \pi$ form factor.  For the $B \to \eta^{(')}$ form factors,
some theoretical studies have been done by assuming the standard quark content
of $\eta^{(\prime)}$ mesons.  Under this assumption, the
$B \to \eta^{(\prime)}$ form factors can be determined by a phenomenological
approach using the semileptonic decays $B \to \eta^{(')} l \nu$ \cite{KimYang},
by using the LCSR model \cite{Aliev}, or by applying the isospin
symmetry to the $B \to \pi $ form factor \cite{Cheng}.
Presently these approaches can give consistent results.
In particular, in Ref. \cite{KimYang} the branching ratios (BRs) for
$B^{\pm} \to \eta l \nu$ and $B^{\pm} \to \eta' l \nu$ are estimated to be
\begin{eqnarray}
{\mathcal B}( B^{\pm} \to \eta l \nu)
&=& (4.32 \pm 0.83) \times 10^{-5},  \nonumber \\
{\mathcal B}( B^{\pm} \to \eta' l \nu)
&=& (2.10 \pm 0.40) \times 10^{-5}.
\label{BRetalnu}
\end{eqnarray}

For last several years the experimental results of unexpectedly large
BRs for $B \to \eta^\prime K$ decays have drawn
a lot of theoretical attentions.
The observed BRs for $B^{\pm} \to \eta' K^{\pm}$ in three different
experiments are \cite{CLEO2, BELLE2, BABAR2}
\begin{eqnarray}
{\mathcal B}( B^{\pm} \to \eta' K^{\pm})
&=& ( 80^{+10}_{-9} \pm 7 ) \times 10^{-6}  ~~[{\rm CLEO}], \nonumber \\
\mbox{} &=& ( 77.9^{+6.2+9.3}_{-5.9-8.7}) \times 10^{-6}
 ~~[{\rm BELLE}],  \nonumber \\
\mbox{} &=& ( 67 \pm 5 \pm 5 ) \times 10^{-6}  ~~[{\rm BABAR}].
\end{eqnarray}
Many theoretical efforts have been made to explain the large BRs:
for instance, approaches using
the anomalous $g{\mbox -}g{\mbox -}\eta^\prime$ coupling
\cite{Soni, Kagan, Kou1, akoy},
high charm content in $\eta^\prime$ \cite{charm, Ko},
the spectator hard scattering mechanism \cite{Du, Yang},
the perturbative QCD approach \cite{pQCD} and
approaches to invoke new physics \cite{dko, Kundu, Kou2, Xiao}.

In Ref. \cite{Neubert} Beneke and Neubert (BN) have tried
to explain the large BRs for $B \to \eta^\prime K$ decays through
the property of the flavor-singlet component of the $\eta^\prime$
meson as well as large weak annihilation effects in the framework
of QCD factorization.  In this approach
it has been suggested that the form factors for the $B \to
\eta^{(\prime)}$ transition may have an additional contribution
through a singlet mechanism, where the
flavor-singlet meson states are produced via the two-gluon emission
in a decaying $B$ meson.
However, since it is unknown how large this new
flavor-singlet contribution ($F^{\rm singlet}_+$) to the $B \to
\eta^{(\prime)}$ form factors is, the uncertainty in
$F^{\rm singlet}_+$ would prevent accurate theoretical estimates of
the BRs for $B \to \eta^{(\prime)} K$.  Indeed, it has been found \cite{Neubert}
that the qualitative pattern of the BRs for $B\to \eta^{(\prime)}
K^{(*)}$ decays can be accounted for in their approach, but
{\it within large uncertainties} mainly coming from
the weak annihilation, and
the strange quark mass and the unknown two-gluon contribution
$F^{\rm singlet}_+$ to the $B\to \eta^{(\prime)}$ form factors.
Therefore, in order to improve the theoretical estimations it is
essential to develop specific methods for estimating
those primary sources of the large uncertainties as accurately as possible.
We note that $B \to \eta^{(\prime)} l \nu$ decays
can be the ideal process to investigate the $B\to \eta^{(\prime)}$ transition
form factors.

In this work we study $B \to \eta^{(\prime)} l \nu$ decays, keeping in mind
the possible effect of the additional flavor-singlet term on these decays.
Our goal is two-fold.  First, we try to present certain phenomenological
methods to determine the flavor-singlet contribution $F^{\rm singlet}_+$
to the $B\to \eta^{(\prime)}$ form factors with a reasonable accuracy,
which is one of the main sources of the large uncertainties involved
in BN's approach.
Secondly, we investigate how large the flavor-singlet term
$F^{\rm singlet}_+$ can affect the predicted BRs for
$B \to \eta^{(\prime)} l \nu$ decays, in comparison with the results
previously presented in literature.
For this aim, we calculate the BRs for $B \to \eta^{(\prime)} l \nu$ decays
including the flavor-singlet effect and compare the result with those
previously given in literature.

The physical states of $\eta$ and $\eta^\prime$ mesons
are described as admixtures of the flavor
octet state $|\eta_8 \rangle = (|u\bar{u}\rangle + |d
\bar{d}\rangle -2 |s \bar{s} )/\sqrt{6}$ and the flavor singlet state
$|\eta_0 \rangle = (|u\bar{u}\rangle + |d \bar{d}\rangle
+|s\bar{s} \rangle)/\sqrt{3}$.  An important clue to the solution
of the $B\to \eta^{(\prime)}$ anomaly may be obtained by examining
the flavor-singlet property of the $\eta^{(\prime)}$ mesons. The
divergence of the axial-vector current is given by
\be
\partial_\mu (\bar{q}\gamma^\mu \gamma_5 q) =
2im_q \bar{q}\gamma_5 q - \frac{\alpha_s}{4\pi} G_{\mu\nu}^a
\tilde{G}^{a,\mu\nu} ~,
\ee
where the last term of the right-handed side is the QCD anomaly term,
and $G^a_{\mu\nu}$ and
$\tilde{G}^{a,\mu\nu}$ are the gluonic field strength tensor and
its dual, respectively. The anomalous contribution to the
divergence of the axial-vector current does not appear in the
flavor-octet state. On the contrary, the flavor-singlet state has
the QCD anomaly contribution and the large mass of the
$\eta^\prime$ meson as compared to other light pseudoscalar
mesons can be explained by this anomaly contribution.

The transition amplitude for the semileptonic decays of a $B$ meson
to an  $\eta^{(\prime)}$ meson can be written as
\be
{\cal M}(B\to \eta^{(\prime)} l \nu ) =
\frac{G_F}{\sqrt{2}} V_{ub} \bar{l} \gamma^\mu (1-\gamma_5) \nu
\langle \eta^{(\prime)}(p_{\eta^{(\prime)}})
 | \bar{u} \gamma_\mu ( 1 - \gamma_5 ) b
| B(p_B) \rangle ~.
\ee
The hadronic matrix element can be parameterized as
\be
\langle \eta^{(\prime)}(p_{\eta^{(\prime)}})
 | \bar{u} \gamma_\mu ( 1 - \gamma_5 ) b
| B(p_B) \rangle = F^{B\to \eta^{(\prime)}}_+ (q^2)
 (p_B + p_{\eta^{(\prime)}})_\mu
+F^{B\to \eta^{(\prime)}}_-  (q^2) (p_B-p_{\eta^{(\prime)}})_\mu ~,
\ee
where $q_{\mu}= (p_B - p_{\eta^{(\prime)}})_{\mu}$.
The differential decay width is given by
\be
\frac{d \Gamma (B \to
\eta^{(\prime)} l \nu)}{d q^2 d \cos \theta} = |V_{ub}|^2
\frac{G_F^2 p_{\eta^{(\prime)}}^3 }{32 \pi^3} \sin^2 \theta
|F^{B\to \eta^{(\prime)}}_+ (q^2)|^2 ~,
\ee
where the mass of the produced lepton has been ignored.
Note that the differential decay
width depends only on one form factor $F^{B\to \eta^{(\prime)}}_+
(q^2)$. Here $\theta$ is the angle between the charged lepton direction
in the virtual $W$ rest frame and the direction of the virtual $W$.

If charm contents and gluonic admixtures of $\eta^{(\prime)}$
mesons are ignored, $\eta^{(\prime)}$ mesons can be related to the
flavor states, $| \eta_{ud} \rangle$ and $ | \eta_s \rangle$ as
\bea
&&|\eta \rangle = \cos \phi | \eta_{ud} \rangle
                - \sin \phi | \eta_s \rangle ~, \no
&&|\eta^\prime  \rangle = \sin \phi | \eta_{ud} \rangle + \cos
\phi | \eta_s \rangle ~,
\label{etaetaprime}
\eea
where $|\eta_{ud} \rangle = \frac{1}{\sqrt{2}} | u\bar{u}+d\bar{d}
\rangle$ and $ |\eta_s \rangle = | s\bar{s} \rangle$.
The best fit value of the mixing angle
$\phi$ is $39.3^\circ \pm 1.0^\circ$ \cite{Feldmann}.
{}From Eq.~(\ref{etaetaprime}), the decay constants are related as
\cite{Feldmann}
\bea
&&f_\eta^{ud} = f_{ud} \cos \phi,
~~~ f_\eta^s = - f_s \sin \phi, \nonumber \\
&&f_{\eta^\prime}^{ud} = f_{ud} \sin \phi,
~~~ f_{\eta^\prime}^s =  f_s \cos \phi,
\label{fetaetaprime}
\eea
where $f_{ud}$ and $f_s$ are the decay constants obtained
from the $\eta_{ud}$ and $\eta_s$ components of the wave functions,
respectively.
Considering a first order correction due to the flavor symmetry
breaking, they can be given by $f_{ud} = f_\pi$ and
$f_s=\sqrt{2f_K^2 - f_\pi^2}$, respectively \cite{Feldmann}.
The phenomenological study about these decay constants yields
\be
f_{ud} = (1.07\pm 0.02) f_\pi, ~~ f_s = (1.34 \pm 0.06) f_\pi.
\label{fudspi}
\ee

\begin{figure}
    \centerline{ \DESepsf(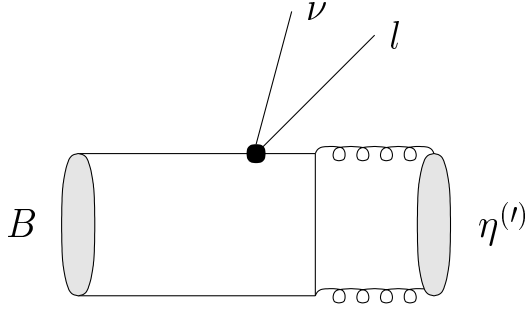 width 7cm)}
    \smallskip
    \caption{\label{fig:fig1} Leading power two-gluon contribution
    to the $B \to \eta^{(\prime)}$ transition form factor. }
\end{figure}

However, for the $B\to \eta^{(\prime)}$ and $B\to \pi$ transition form
factors, the simple relation such as Eq. (\ref{fudspi}) may not hold,
because $\eta^{(\prime)}$ mesons have the flavor-singlet meson state
produced via the two-gluon emission from the light spectator quark
\cite{Neubert}, as shown in Fig.~\ref{fig:fig1}.
This diagram gives a leading power correction to the $B\to
\eta^{(\prime)}$ form factors.
We parameterize the new two-gluon contribution, which is proportional to
the flavor-singlet decay constant,
as $F_+^{\rm singlet}$.  {}From now on, we will call this $F_+^{\rm singlet}$
the flavor-singlet form factor.  Including this flavor-singlet
contribution $F_+^{\rm singlet}$, the hadronic form factors
in the $B \to \eta^{(\prime)}$ transition can be expressed as \cite{Neubert}
\bea
F_+^{B\to \eta}(0)
\areq \frac{1}{\sqrt{2}} \frac{ f_\eta^{ud}}{f_\pi} F_+^{B\to \pi}(0) +\frac{
(\sqrt{2} f_\eta^{ud} + f_\eta^s)} {\sqrt{3}f_\pi }
 F_+^{\rm singlet}(0) ~,
\no
F_+^{B\to \eta^{\prime}}(0) \areq \frac{1}{\sqrt{2}}
\frac{f_{\eta^\prime}^{ud}}{f_\pi}
F_+^{B\to \pi}(0)
+ \frac{  (\sqrt{2} f_{\eta^\prime}^{ud} + f_{\eta^\prime}^s)}
{\sqrt{3} f_\pi}
 F_+^{\rm singlet}(0) ~,
\eea
where $\pi$ in the superscript denotes the charged pion.

In order to determine the flavor-singlet form factor
$F_+^{\rm singlet}(0)$, we present three observables \cite{KimYang}
which can be measured in experiment.
These observables, $R_1$, $R_2$ and $R_3$, are defined by
ratios of the differential decay rates of the relevant modes measured
at maximum recoil point ($q^2=0$) as follows:
\begin{eqnarray}
&&R_{1 (2)} \equiv
\frac{d \Gamma (B^- \to \eta^{(\prime)} l \nu )/d q^2 }
{d \Gamma (B^- \to \pi^0 l \nu )/d q^2 } \Bigg|_{q^2 =0}
\nonumber \\
&&\hspace{0.85cm}
= \frac{(m_B^2 - m_{\eta^{(\prime)}}^2)^3}{(m_B^2 - m_\pi^2)^3}
\Bigg| \frac{f_{\eta^{(\prime)}}^{ud}}{f_\pi} +
\sqrt{2} \tilde F
\frac{(\sqrt{2}f_{\eta^{(\prime)}}^{ud} + f_{\eta^{(\prime)}}^s)}
{\sqrt{3}f_\pi} \Bigg|^2,
\nonumber \\
&&R_{3} \equiv
\frac{d \Gamma (B^- \to \eta^\prime l \nu )/d q^2 }
{d \Gamma (B^- \to \eta l \nu )/d q^2 } \Bigg|_{q^2 =0}
\nonumber \\
&&\hspace{0.6cm}
= \frac{(m_B^2 - m_{\eta^\prime}^2)^3}{(m_B^2 - m_\eta^2)^3}
\Bigg| \frac{\sqrt{3} f_{\eta^\prime}^{ud}
+ \sqrt{2} \tilde F (\sqrt{2}f_{\eta^\prime}^{ud}
+ f_{\eta^\prime}^s) }
{\sqrt{3} f_{\eta}^{ud}
+ \sqrt{2} \tilde F (\sqrt{2}f_{\eta}^{ud}
+ f_{\eta}^s) }
 \Bigg|^2 ,
\label{RRR}
\end{eqnarray}
where $\tilde F \equiv F_+^{\rm singlet}(0) / F_+^{B\to \pi} (0)$.
The differential decay rates $d \Gamma /d q^2$ for the semileptonic
$B$ decays at $q^2 =0$ can be experimentally measured: for example,
the $d \Gamma /d q^2$ distribution, including at $q^2 =0$, for
$\bar B \to D^* l \bar \nu$ decays has been measured by the CLEO
Collaboration \cite{CLEO3}.
We note that the theoretical estimates of $R_1$, $R_2$ and $R_3$
depend only on the ratio, $\tilde F$, of the form factors
$F^{\rm singlet}_+ (0)$ and $F^{B \to \pi}_+ (0)$, besides the relevant
meson masses and the decay constants, $f_{\pi}$ and $f_{\eta^{(\prime)}}$.
In the above observables, the CKM matrix element $|V_{ub}|$ does not appear
due to cancellation between the denominator and the numerator, so the large
uncertainty involved in $|V_{ub}|$ can be avoided.

\begin{table}
\caption{Parameters of the $B \to \pi $ transition form factor in
the lattice I, II, and LCSR  models.}
\smallskip
\begin{tabular}{|c|c|c|c|c|}
\hline
& Lattice~  I &
 Lattice~  II &
~~~ LCSR ~~~&~Average Value~\\
\hline
~${F_+^{B\to \pi}(0)}$~ & ~${0.26\pm 0.05\pm 0.04}$~&
~${0.28\pm 0.06 \pm 0.05}$~& ~$0.28\pm 0.05$~ &~$0.27\pm 0.04$~\\
~$c_B$~&~ $ 0.42\pm 0.13\pm 0.04$~&~ $ 0.51 \pm 0.08 \pm 0.01$~ &
~$0.41\pm 0.12$~&~$0.47\pm 0.06$~\\
~$\alpha_B$~ &~ $0.40 \pm 0.15 \pm 0.09$~ &~ $ 0.45\pm 0.17^{+0.06}_{-0.13}$
~&~$0.32^{+0.21}_{-0.07}~$&~$0.39\pm 0.11$~\\
\hline
\end{tabular}
\end{table}

The $B\to \pi$ form factor $F^{B\to \pi}_+$ has been studied
in many models: for instance, the lattice calculation \cite{lattice}
and the LCSR \cite{lightcone1}, and so on.
The form factor for the $B\to \pi $ transition
can be expressed \cite{BK} as \be F_+^{B\to \pi}(q^2) =
\frac{F_+^{B\to \pi} (0)}{(1-\tilde{q}^2)(1-\alpha_B \tilde{q}^2)}
~, \label{form} \ee where $\tilde{q}^2 = q^2 /m_{B^\ast}^2$,
$F_+^{B\to \pi}(0)=c_B (1 -\alpha_B)$ and $m_{B^\ast}$ is the mass
of $B^\ast$ meson. It is well known that Eq.~(\ref{form})
satisfies most of the known constraints on the form factor, such
as the heavy quark scaling laws predicted by the HQET in the zero
recoil region $(q^2 \to q_{\rm max}^2)$ \cite{HQETscaling} and by
the large energy effective theory in the large recoil region
$(q^2\to 0)$ \cite{LEET}. To extrapolate parameters of the
relevant form factor, we adopt two different lattice calculation
models (Lattice I and II) \cite{lattice, UKQCD} and the LCSR model
\cite{lightcone1}. Specific values of the $B\to \pi$ form factor
for each model are shown in Table I, where the last column is the
weighted-average values of those three models. Those values are in
good agreement with each other.

\begin{figure}
    \centerline{ \DESepsf(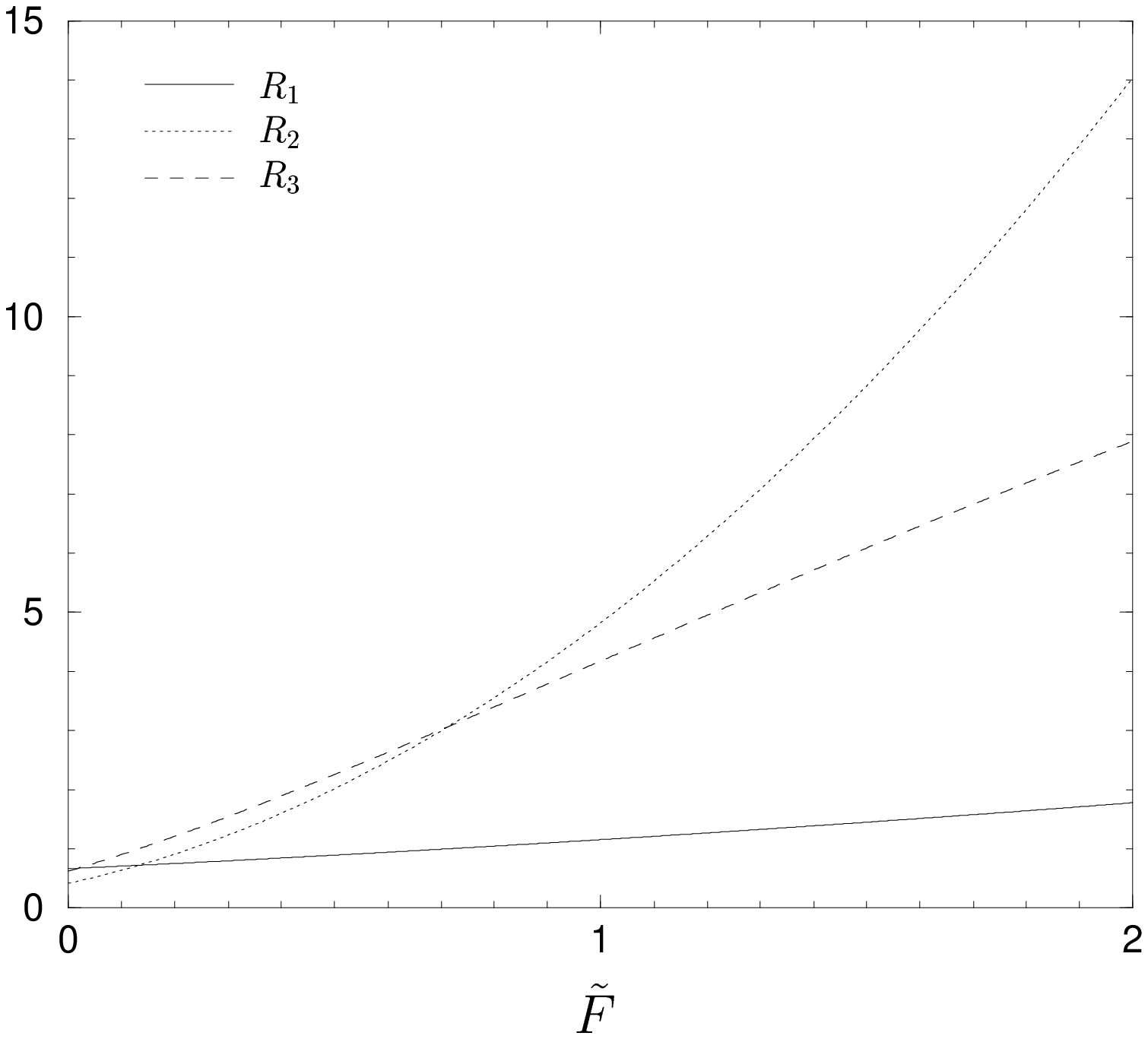 width 9cm)}
    \smallskip
    \caption{\label{fig:fig2} The predicted $R_1$, $R_2$ and
     $R_3$ versus the ratio $\tilde F \equiv F_+^{\rm singlet}(0) /
     F_+^{B\to \pi} (0)$. }
\end{figure}

Using the values of $F^{B \to \pi}_+ (0)$ given in Table I, the
observables $R_1$, $R_2$ and $R_3$ are estimated for
$F^{\rm singlet}_+ (0) = 0, ~0.2, ~0.4, ~0.6$ in Tables II, III
and IV.
For the values of the relevant decay constants, we have used
the relations and the central values given in
Eqs. (\ref{fetaetaprime}) and (\ref{fudspi}).
In the case of $F^{\rm singlet}_+ (0) = 0$ there is no
uncertainty from the form factors, because $R_1$, $R_2$
and $R_3$ are dependent only on the ratio
$\tilde F \equiv F_+^{\rm singlet}(0) / F_+^{B\to \pi} (0)$.
It is interesting to observe that $R_2$ and $R_3$ are very
sensitive to $F_+^{\rm singlet}(0)$.  For instance, for
$F_+^{\rm singlet}(0) =0.2$, $R_2$ and $R_3$ are
$3.14 \pm 0.70$ and $3.12 \pm 0.45$, respectively, while for
$F_+^{\rm singlet}(0) =0$, $R_2 =0.42$ and $R_3 =0.63$.
In Fig. \ref{fig:fig2}, we show the predicted $R_1$, $R_2$
and $R_3$ as a function of the ratio $\tilde F$.  Again it is
clearly seen that $R_2$ and $R_3$ change sensitively as
the ratio $\tilde F$ varies, while $R_1$ is rather
insensitive to $\tilde F$.
Thus, by measuring $R_2$ and/or $R_3$ in experiment, one
can determine the value of $F_+^{\rm singlet}(0)$.
Once the value of the form factor $F_+^{B \to \pi}(0)$ becomes
more accurately known through both theoretical and experimental
studies, the flavor-singlet form factor $F_+^{\rm singlet}(0)$
can be determined more precisely.
We should emphasize that in this method there is no ambiguity
related to the unknown $q^2$ dependence of $F_+^{\rm singlet}$,
which can cause a large uncertainty to theoretical calculations
of quantities of interest, such as the BRs for
$B \to \eta^{(\prime)} l \nu$ (See below).

\begin{table}
\caption{${\cal R}_1$  for $F_+^{\rm singlet}(0) =
0,~0.2,~0.4,~0.6$.}
\smallskip
\begin{tabular}{|c|cccc|}\hline
 &\multicolumn{4}{|c|}{~~${\cal R}_1$ }
\\ \hline
$F_+^{\rm singlet}(0)$ &~0~ & ~0.2~ & ~0.4~&~0.6~\\
\hline
~Lattice Model I~
&$ 0.67 $
&$ 1.03^{+0.12}_{-0.07}$
&$ 1.47^{+0.30}_{-0.18}$
&$ 2.00^{+0.54}_{-0.30}$
\\
~Lattice Model II~
&$ 0.67$
&$ 1.00^{+0.15}_{-0.08}$
&$ 1.41^{+0.37}_{-0.18}$
&$ 1.88^{+0.66}_{-0.32}$
\\
~LCSR~
&$ 0.67$
&$ 1.00^{+0.08}_{-0.06}$
&$ 1.41^{+0.20}_{-0.13}$
&$ 1.88^{+0.35}_{-0.22}$
\\
\hline
~Average~
&$0.67$&$1.01\pm 0.06$&$1.42\pm 0.15$&$1.91\pm 0.27$
\\
\hline
\end{tabular}
\end{table}

\begin{table}
\caption{${\cal R}_2$ for $F_+^{\rm singlet}(0) =
0,~0.2,~0.4,~0.6$.}
\smallskip
\begin{tabular}{|c|cccc|}\hline
&\multicolumn{4}{|c|}{~~${\cal R}_2$ }
\\ \hline
$F_+^{\rm singlet}(0)$ &~0~ & ~0.2~ & ~0.4~&~0.6~\\
\hline
~Lattice Model I~
&$ 0.42$
&$ 3.38^{+1.44}_{-0.77}$
&$ 9.18^{+4.85}_{-2.51}$
&$17.85^{+10.23}_{-5.22}$
\\
~Lattice Model II~
&$ 0.42$
&$ 3.07^{+1.75}_{-0.80}$
&$ 8.19^{+5.86}_{-2.58}$
&$15.75^{+12.32}_{-5.32}$
\\
~LCSR~
&$ 0.42$
&$ 3.07^{+0.90}_{-0.56}$
&$ 8.19^{+2.99}_{-1.81}$
&$15.75^{+ 6.26}_{-3.74}$
\\
\hline
~Average~
&$0.42$&$3.14\pm 0.70$&$8.42\pm 2.33$&$16.23\pm 4.90$\\
\hline
\end{tabular}
\end{table}

\begin{table}
\caption{${\cal R}_{3}$ for $F_+^{\rm singlet}(0) =
0,~0.2,~0.4,~0.6$.}
\smallskip
\begin{tabular}{|c|cccc|}\hline
&\multicolumn{4}{|c|}{~~${\cal R}_{3} $ }
\\ \hline
$F_+^{\rm singlet}(0)$ &~0~ & ~0.2~ & ~0.4~&~0.6~\\
\hline
~Lattice Model I~
&$ 0.63$
&$ 3.28^{+0.90}_{-0.55}$
&$ 6.23^{+1.66}_{-1.09}$
&$ 8.93^{+ 2.13}_{-1.48}$
\\
~Lattice Model II~
&$ 0.63$
&$ 3.07^{+1.10}_{-0.60}$
&$ 5.82^{+2.07}_{-1.21}$
&$ 8.38^{+ 2.68}_{-1.68}$
\\
~LCSR~
&$ 0.63$
&$ 3.07^{+0.60}_{-0.41}$
&$ 5.82^{+1.15}_{-0.82}$
&$ 8.38^{+ 1.51}_{-1.13}$
\\
\hline
~Average~
&$0.63$&$3.12\pm 0.45$&$5.93\pm 0.86$&$ 8.53\pm 1.12$\\
\hline
\end{tabular}
\end{table}

Now we examine the possible contribution of the flavor-singlet term
to the BRs for $B \to \eta^{(\prime)} l \nu$.
In order to estimate the BRs, one needs to know the $q^2$ dependence
of the flavor-singlet form factor $F_+^{\rm singlet}$.  Since it is
completely unknown, for illustration we assume that
$F_+^{\rm singlet}(q^2)$ has the same $q^2$ dependence as
$F_+^{B\to \pi}(q^2)$.  Since the $B^*$ pole contribution to the
form factor could be different for the pion and the singlet state,
this assumption could cause an uncertainty to the estimate.

\begin{figure}
    \centerline{ \DESepsf(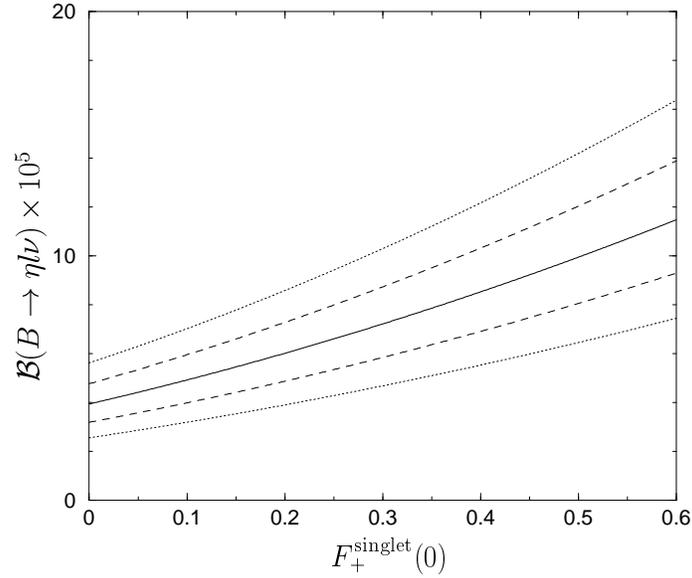 width 9cm)}
    \smallskip
    \caption{\label{fig:fig3} BR (in $10^{-5}$) of
    $B^- \to \eta l^- \bar \nu$ versus the flavor-singlet contribution
    $F_+^{\rm singlet}(0)$. The solid line is for $|V_{ub}|
    = 3.6 \times 10^{-3}$ which is the central value of the average,
    $|V_{ub}| = (3.6 \pm 0.7) \times 10^{-3}$,
    of the LEP and the CLEO data.  The region between the dotted lines
    corresponds to the uncertainty arising from $1 \sigma$ error in
    $|V_{ub}|$.
    For the dashes lines, we have assumed 10 \% error in $|V_{ub}|$.}
\end{figure}

\begin{figure}
    \centerline{ \DESepsf(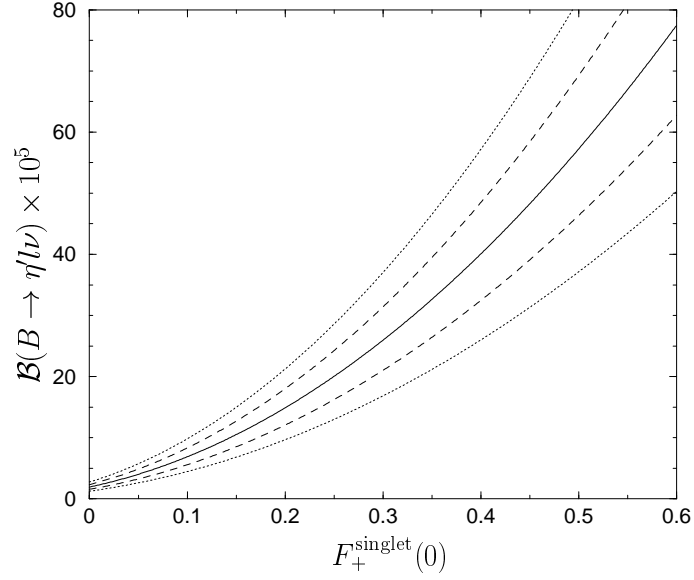 width 9cm)}
    \smallskip
    \caption{\label{fig:fig4} BR (in $10^{-5}$) of
    $B^- \to \eta^\prime l^- \bar \nu$ versus the flavor-singlet contribution
    $F_+^{\rm singlet}(0)$. The definition of the lines are the same as in
    Fig. 4.}
\end{figure}

In the numerical calculation, we use the average value of the CLEO
and LEP data: $|V_{ub}|= (3.6\pm 0.7)\times 10^{-3}$, which
does not include the preliminary results from BELLE and BABAR.
The estimated BRs of the decays $B^- \to \eta l^- \bar \nu$ and
$B^- \to \eta' l^- \bar \nu$ as a function of $F_+^{\rm singlet}(0)$
are shown in Figs. \ref{fig:fig3} and \ref{fig:fig4}, respectively.
In the figures, the solid line denotes the case of using
$|V_{ub}| = 3.6 \times 10^{-3}$, which is the central value of the
average of the CLEO and LEP data.
The estimated BRs are
\bea
{\cal B}(B^- \to \eta l^- \bar \nu) &=& (4.00 \pm 0.99) \times 10^{-5},
\nonumber \\
{\cal B}(B^- \to \eta' l^- \bar \nu) &=& (1.95 \pm 0.48) \times
10^{-5},
\eea
for $F_+^{\rm singlet}(0) =0$.
These values are in good agreement with those in Eq. (\ref{BRetalnu}).
The dotted lines denote the case of allowing $1 \sigma$ error in the
CLEO and LEP data: i.e., the upper and lower dotted lines are for
$|V_{ub}| = 4.3 \times 10^{-3}$ and $2.9 \times 10^{-3}$,
respectively.
We also present the case of assuming 10\% error in
the value of $|V_{ub}|$ as the dashed lines: i.e., the upper and
lower dashed lines are for $|V_{ub}| = 3.96 \times 10^{-3}$ and
$3.24 \times 10^{-3}$, respectively.
For some representative values of $F_+^{\rm singlet}(0)$, the BRs
of $B\to \eta^{(\prime)} l \nu$ estimated in each model are shown
in Tables V and VI.
We see that the estimated BR of $B\to \eta' l \nu$ is particularly
sensitive to the effect of $F_+^{\rm singlet}(0)$: for example,
the BR increases from $(1.95 \pm 0.48)$ to $(14.99 \pm 3.70)$ as
$F_+^{\rm singlet}(0)$ varies from 0 to 0.2.
This feature can be easily understood by the fact that the
flavor-singlet component of $\eta'$ occupies more contribution
than that of $\eta$.

Assuming the $q^2$ dependence of $F_+^{\rm singlet}$ is correct,
Figs.~\ref{fig:fig3} and \ref{fig:fig4} could provide another way
of determining the value of $F_+^{\rm singlet}(0)$.
That is, one can measure the BR of $B^- \to \eta l^- \bar \nu$ and/or
$B^- \to \eta' l^- \bar \nu$ in experiment and then could determine
$F_+^{\rm singlet}(0)$ by using Fig.~\ref{fig:fig3} and/or
\ref{fig:fig4}.
For that purpose, the process $B^- \to \eta' l^- \bar \nu$ would be
more useful than $B^- \to \eta l^- \bar \nu$, due to the strong
sensitivity of its BR to $F_+^{\rm singlet}(0)$ and the smaller
uncertainty arising from the error in $|V_{ub}|$, as shown in
Fig. \ref{fig:fig4}.

\begin{table}
\caption{BR of $B\to \eta l \nu $ decay for $F_+^{\rm
singlet}(0)=0, ~0.2, ~0.4, ~0.6$.}
\smallskip
\begin{tabular}{|c|cccc|}\hline
 &\multicolumn{4}{|c|}{~~${\cal B}(B\to \eta l \nu)\times 10^5$ }
\\ \hline
$F_+^{\rm singlet}(0)$ &~~~~0~~~~&~0.2~&~0.4~&~0.6~
\\
\hline ~Lattice I~ &$3.68^{+1.57}_{-1.29} $
& $ 5.70^{+2.43}_{-2.00}$
& $ 8.16^{+3.48}_{-2.86}$
& $11.06^{+ 4.72}_{- 3.88}$
\\
~Lattice II~
& $ 4.43^{+1.89}_{-1.55}$
& $ 6.67^{+2.85}_{-2.34}$
& $ 9.37^{+4.00}_{-3.29}$
& $12.52^{+ 5.34}_{- 4.39}$
\\
~LCSR ~
& $ 4.04^{+1.72}_{-1.42}$
& $ 6.08^{+2.59}_{-2.14}$
& $ 8.53^{+ 3.64}_{- 2.99}$
& $11.41^{+ 4.86}_{- 4.01}$
\\
\hline ~Average ~ &$4.00\pm 0.99$ & $ 6.10\pm 1.50$ & $ 8.63\pm
2.13$ & $11.60\pm  2.86$
\\
\hline
\end{tabular}
\end{table}
\begin{table}
\caption{BR of $B \to \eta^\prime l \nu$ decay for $F_+^{\rm
singlet}(0)=0, ~0.2, ~0.4, ~0.6$.}
\smallskip
\begin{tabular}{|c|cccc|}\hline
&\multicolumn{4}{|c|}{~~${\cal B}(B \to \eta^\prime l \nu)\times 10^5$ }
\\ \hline
$F_+^{\rm singlet}(0)$
&~~~~0~~~~&~0.2~&~0.4~&~0.6~\\
\hline
~Lattice I~
& $ 1.80^{+ 0.76}_{- 0.64}$
& $14.57^{+ 6.22}_{- 5.11}$
& $39.63^{+16.91}_{-13.91}$
& $79.96^{+32.84}_{-24.02}$
\\
~Lattice II~
& $ 2.14^{+ 0.92}_{- 0.75}$
& $15.81^{+ 6.75}_{- 5.55}$
& $42.12^{+17.97}_{-14.79}$
& $81.05^{+34.59}_{-28.45}$
\\
~LCSR ~
& $ 1.99^{+ 0.85}_{- 0.21}$
& $14.72^{+ 6.28}_{- 5.17}$
& $39.20^{+16.72}_{-13.76}$
& $75.43^{+32.19}_{-26.67}$
\\
\hline
~Average ~
&$1.95\pm 0.48$ & $14.99\pm 3.70$ & $40.24\pm  9.92$ & $78.69\pm 19.15$\\
\hline
\end{tabular}
\end{table}

To avoid the large uncertainty in the parameter $|V_{ub}|$,
the ratios of the BRs, ${\cal B}(B^- \to \eta^{(\prime)}l \nu )$
and ${\cal B}(B^- \to \pi^0 l \nu)$, have been suggested
in Ref. \cite{KimYang}.
They are now modified after including the flavor-singlet form factor as
\bea
{\cal R}_{\eta}
~&\hspace{-2.35mm}\equiv&\hspace{-2.35mm}~
\frac{{\cal B}(B^- \to \eta l \nu)}{{\cal B}(B^- \to \pi^0 l \nu)}
= \left| \frac{f_\eta^{ud}}{f_\pi}
+ \sqrt{2} \tilde F
\frac{  (\sqrt{2} f_\eta^{ud} + f_\eta^{s})}{\sqrt{3}f_\pi } \right|^2
\no
&& \times \frac{ \int^{(m_B-m_\eta)^2}_{0} d q^2
|F_+^{B\to \pi} (q^2)|^2 [(m_B^2+m_\eta^2-q^2)^2 - 4m_B^2 m_\eta^2 ]^{3/2}}
{ \int^{(m_B-m_\pi)^2}_{0} d q^2
|F_+^{B\to \pi} (q^2)|^2 [(m_B^2+m_\pi^2-q^2)^2 - 4m_B^2 m_\pi^2 ]^{3/2}} ~~,
\no
{\cal R}_{\eta^\prime}
~&\hspace{-2.35mm}\equiv&\hspace{-2.35mm}~
\frac{{\cal B}(B^- \to \eta^\prime l \nu)}{{\cal B}(B^- \to \pi^0 l \nu)}
= \left| \frac{f_{\eta^\prime}^{ud}}{f_\pi}
+ \sqrt{2} \tilde F
\frac{  (\sqrt{2} f_{\eta^\prime}^{ud} + f_{\eta^\prime}^{s})}{\sqrt{3}f_\pi }
 \right|^2
\no
&& \times \frac{ \int^{(m_B-m_{\eta^\prime})^2}_{0} d q^2
|F_+^{B\to \pi} (q^2)|^2 [(m_B^2+m_{\eta^\prime}^2-q^2)^2
- 4m_B^2 m_{\eta^\prime}^2 ]^{3/2}}
{ \int^{(m_B-m_\pi)^2}_{0} d q^2
|F_+^{B\to \pi} (q^2)|^2 [(m_B^2+m_\pi^2-q^2)^2 - 4m_B^2 m_\pi^2 ]^{3/2}} ~~,
\eea
where the dependence on $|V_{ub}|$ cancels out between the numerator and
the denominator.

\begin{figure}
    \centerline{ \DESepsf(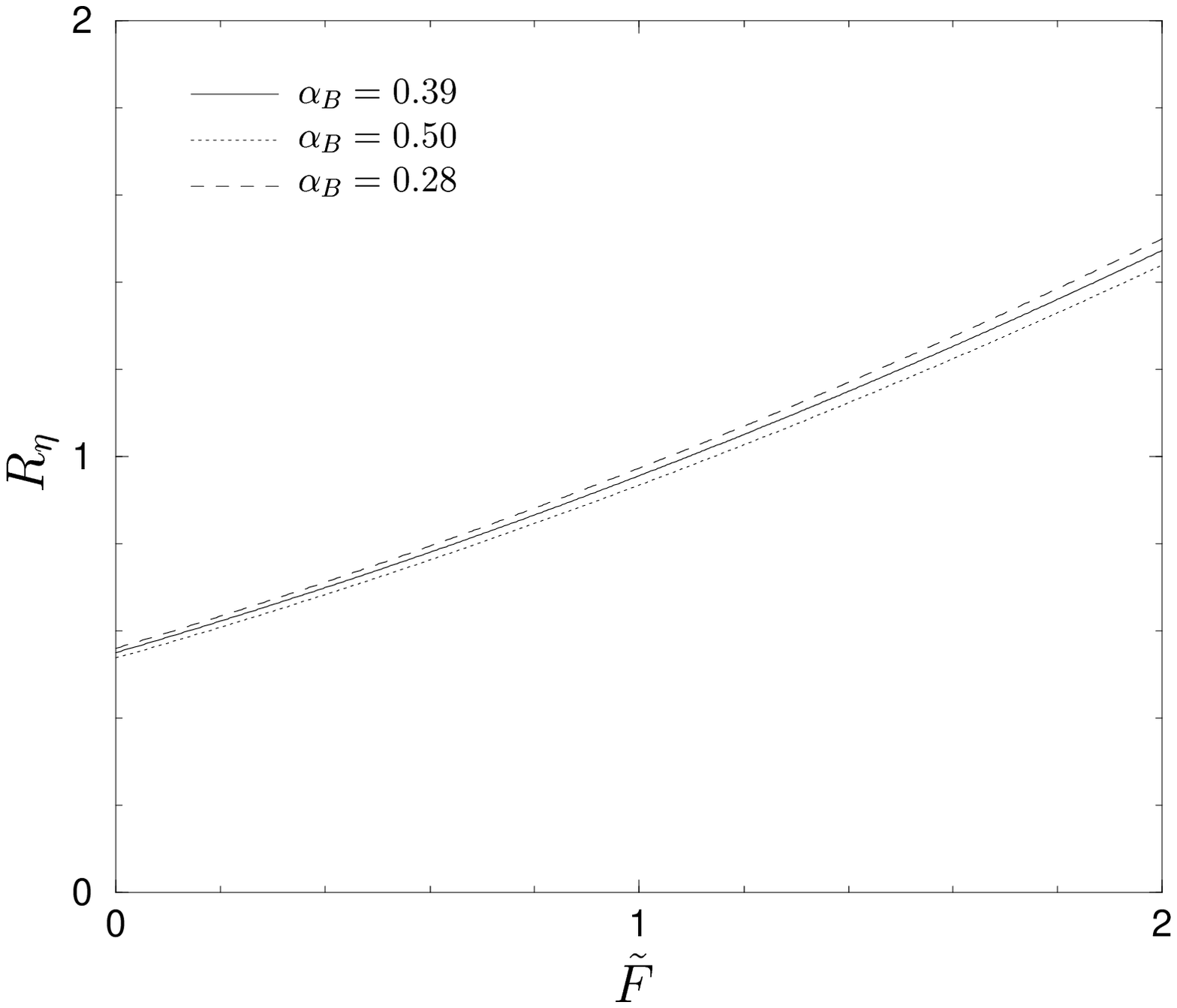 width 9cm)}
    \smallskip
    \caption{\label{fig:fig5} $R_{\eta}$ versus $\tilde F \equiv
    F_+^{\rm singlet}(0) / F^{B \to \pi}_+(0)$.
    The solid, the dotted, and the dashed lines
    correspond to $\alpha_B =0.39, ~0.50, ~0.28$, respectively. }
\end{figure}

\begin{figure}
    \centerline{ \DESepsf(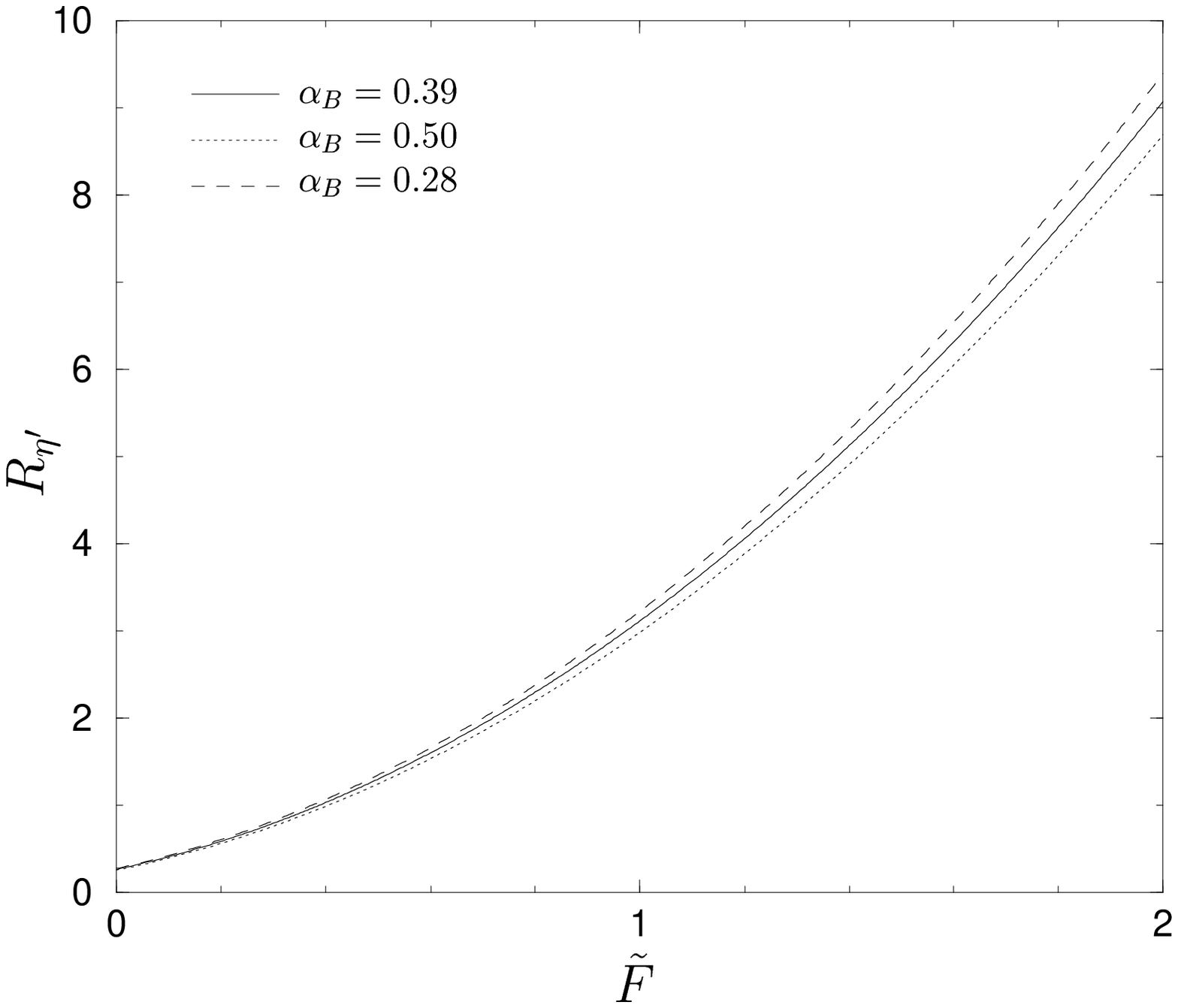 width 9cm)}
    \smallskip
    \caption{\label{fig:fig6} $R_{\eta^\prime}$ versus $\tilde F \equiv
    F_+^{\rm singlet}(0) / F^{B \to \pi}_+(0)$.
    The solid, the dotted, and the dashed lines correspond to
    $\alpha_B =0.39, ~0.50, ~0.28$, respectively. }
\end{figure}

Tables VII and VIII show our predictions of ${\cal R}_\eta$
and ${\cal R}_{\eta^\prime}$, respectively, in Lattice Model
I, II and LCSR for $F_+^{\rm singlet}(0) = 0, 0.2, 0.4, 0.6$,
where we have used the average value of
$F_+^{B \to \pi}(q^2)$ with the $1\sigma$ error for
$F_+^{B\to \pi}(0)$ and $\alpha_B$ given in Table I.
In Figs.~\ref{fig:fig5} and \ref{fig:fig6}, we present the
predicted ${\cal R}_\eta$ and ${\cal R}_{\eta^\prime}$
as a function of $\tilde F \equiv F_+^{\rm singlet}(0)
/ F_+^{B \to \pi}(0)$.
Here the uncertainty in $\alpha_B$ shown in Table I has been
also considered, and denoted as the dotted line ($\alpha_B = 0.50$)
and the dashed line $(\alpha_B = 0.28)$.
The dependence of ${\cal R}_{\eta^{(\prime)}}$ on the uncertainty
in $\alpha_B$ is rather weak.
As expected, ${\cal R}_{\eta'}$ are particularly sensitive
to $\tilde F$.
Again, assuming the relation (\ref{form}) holds for the singlet
state as well, Figs.~\ref{fig:fig5} and \ref{fig:fig6} could
provide another alternative way of determining the value of
$F_+^{\rm singlet}(0)$, without suffering the large uncertainty
in $|V_{ub}|$: i.e., one could determine the value of
$F_+^{\rm singlet}$ by measuring ${\cal R}_{\eta}$ and/or
${\cal R}_{\eta^\prime}$ and using Fig.~\ref{fig:fig5} and/or
\ref{fig:fig6}.

\begin{table}
\caption{${\cal R}_{\eta}$  for $F_+^{\rm singlet}(0) =
0,~0.2,~0.4,~0.6$.}
\smallskip
\begin{tabular}{|c|cccc|}\hline
 &\multicolumn{4}{|c|}{~~${\cal R}_\eta $ }
\\ \hline
$F_+^{\rm singlet}(0)$ &~0~ & ~0.2~ & ~0.4~&~0.6~\\
\hline
~Lattice Model I~
&$ 0.55 \pm 0.01$
&$ 0.85^{+0.10}_{-0.07}$
&$ 1.22^{+0.25}_{-0.16}$
&$ 1.65^{+0.45}_{-0.24}$
\\
~Lattice Model II~
&$ 0.54^{+0.02}_{-0.03}$
&$ 0.82^{+0.13}_{-0.09}$
&$ 1.15^{+0.31}_{-0.17}$
&$ 1.54^{+0.54}_{-0.28}$
\\
~LCSR~
&$ 0.56^{+0.01}_{-0.02}$
&$ 0.84^{+0.07}_{-0.06}$
&$ 1.18^{+0.16}_{-0.12}$
&$ 1.57^{+0.29}_{-0.20}$
\\
\hline
~Average~
&$0.55\pm 0.01$&$0.84\pm 0.05$&$1.19\pm 0.12$&$1.58\pm 0.22$
\\
\hline
\end{tabular}
\end{table}
\begin{table}
\caption{${\cal R}_{\eta^\prime}$ for $F_+^{\rm singlet}(0) =
0,~0.2,~0.4,~0.6$.}
\smallskip
\begin{tabular}{|c|cccc|}\hline
&\multicolumn{4}{|c|}{~~${\cal R}_{\eta^\prime} $ }
\\ \hline
$F_+^{\rm singlet}(0)$ &~0~ & ~0.2~ & ~0.4~&~0.6~\\
\hline
~Lattice Model I~
&$ 0.27^{+0.01}_{-0.02}$
&$ 2.18^{+0.94}_{-0.53}$
&$ 5.92^{+3.14}_{-1.56}$
&$11.49^{+6.61}_{-3.46}$
\\
~Lattice Model II~
&$ 0.26 \pm 0.02$
&$ 1.94^{+1.12}_{-0.55}$
&$ 5.17^{+3.72}_{-1.71}$
&$ 9.95^{+7.82}_{-3.51}$
\\
~LCSR~
&$ 0.23^{+0.01}_{-0.03}$
&$ 2.03^{+0.61}_{-0.40}$
&$ 5.41^{+2.00}_{-1.26}$
&$10.41^{+4.18}_{-0.59}$
\\
\hline
~Average~
&$0.26\pm 0.01$&$2.05\pm 0.47$&$5.49\pm 1.54$&$10.59\pm 3.22$\\
\hline
\end{tabular}
\end{table}

In conclusion, we studied semileptonic decays $B \to
\eta^{(\prime)} l \nu$, considering the flavor singlet
contribution, which arises from the effect of the two-gluon
emission in a decaying $B$ meson.
Using $B^- \to \eta^{(\prime)} l \bar \nu$
(and $B^- \to \pi^0 l \bar \nu$) decays, we demonstrated
how to determine the flavor-singlet form factor
$F_+^{\rm singlet}$ whose unknown value is one of the main
sources of the large uncertainty in theoretical estimates of
the BRs for $B \to \eta' K$ decays in the QCD factorization scheme.
The uncertainty involved in determination of $F_+^{\rm singlet}(0)$
using our method mainly comes from the uncertainty in
$F_+^{B \to \pi}(0)$.  Therefore, as the more accurate value of
$F_+^{B \to \pi}(0)$ becomes available, $F_+^{\rm singlet}(0)$ can
be more precisely determined in forthcoming studies.
We also calculated the BRs for $B^- \to \eta^{(\prime)} l \nu$
and examined how large the effect of $F_+^{\rm singlet}$
on these BRs can be.  Our result shows that the estimated BR for
$B^- \to \eta' l \bar \nu$ strongly depends on the effect of
$F_+^{\rm singlet}$.
\\


\centerline{\bf ACKNOWLEDGEMENTS}
\medskip
\noindent
We thank Y. Kwon for useful discussions.
The work of C.S.K. was supported in part by  CHEP-SRC Program,
in part by Grant No. R02-2003-000-10050-0 from BRP of the KOSEF.
The work of S.O. was supported in part by Grant No. R02-2003-000-10050-0
from BRP of the KOSEF, and by the Japan Society for the
Promotion of Science (JSPS).
The work of C.Y. was supported by Grant No. 2001-042-D00022 of the KRF.

\newpage


\newpage

\end{document}